\RequirePackage{fix-cm}
\documentclass[pdftex,twocolumn,epjc3]{svjour3}
\smartqed  % flush right qed marks, e.g. at end of proof
\usepackage{booktabs}

\usepackage{graphicx}
\graphicspath{{fig/}}
\usepackage{mathptmx} % use Times fonts if available on your TeX system
\usepackage[colorlinks, citecolor=blue, urlcolor=blue, linkcolor=blue, bookmarksdepth=3]{hyperref}
\usepackage{amsmath}% more environments

\usepackage{amssymb}% more symbols
\usepackage{physics}% more commands
\usepackage{bm}% bold math symbol
\usepackage[numbers,sort&compress]{natbib}

\usepackage{siunitx}% typeset units

\journalname{Eur. Phys. J. C}
\begin{document}

\title{First operation of undoped CsI directly coupled with SiPMs at 77 Kelvin}

%\titlerunning{Undoped NaI/CsI at 77 K for NSI detection at the SNS} % if too long for running head

\author{Keyu Ding \and Jing Liu\thanksref{e1} \and Yongjin Yang \and Dmitry Chernyak\thanksref{e2}}

\thankstext{e1}{e-mail: \href{mailto:jing.liu@usd.edu}{jing.liu@usd.edu}}
\thankstext{e2}{now at University of Alabama}
\institute{Department of Physics, University of South Dakota, 414 East Clark Street, Vermillion, SD 57069, USA}

\date{Received: date / Accepted: date}
% The correct dates will be entered by the editor

\maketitle

\begin{abstract}
 The light yield of a small undoped cesium iodide (CsI) crystal directly coupled with two silicon photomultipliers (SiPMs) at about 77~Kelvin was measured to be $43.0 \pm 1.1$~photoelectrons (PE) per keV electron-equivalent (keV$_\text{ee}$) using $X$ and $\gamma$-ray peaks from an $^{241}$Am radioactive source from 18 to 60 keV. The high light yield together with some other technical advantages illustrate the great potential of this novel combination for neutrino and low-mass dark matter detection, particularly at accelerator-based neutrino sources, where random background can be highly suppressed by requiring coincident triggers between SiPMs and beam pulse timing signals. Some potential drawbacks of using cryogenic SiPMs instead of photomultiplier tubes (PMTs) were identified, such as worse energy resolution and optical cross-talks between SiPMs. Their influence to rare-event detection was discussed and possible solutions were provided.
\end{abstract}

\section{Introduction}
Inorganic scintillating detectors are widely used in the detection of dark matter~\cite{dama20, picolon18, dmice17, anais21, cosine21, sabre19, angloher17} and neutrinos~\cite{coherent17} due to their relatively high light yields and easy light readout with PMTs at room temperature. The sensitivity of such a detector can be improved by increasing its target mass and decreasing its energy threshold as more dark matter or neutrino events are expected at lower energies~\cite{freedman74, bar05, coherent17, coherent21}. There are two large limiting factors in decreasing the energy threshold of such a detector~\cite{coherent17}. The first one is the Cherenkov radiation caused by charged particles passing through the quartz window of a PMT~\cite{coherent17}. The second one is the afterglow of the crystal itself after some bright scintillation events~\cite{col15}.

The first limiting factor can be eliminated by replacing PMTs with SiPM arrays, which do not have quartz windows. However, at room temperature, SiPMs exhibit much higher dark count rates (DCR) than PMTs~\cite{sipm}. In order to have a manageable DCR, SiPMs can be operated at cryogenic temperatures~\cite{akiba09, catalanotti15, ost15, igarashi16, aalseth17, giovanetti17}. The cryogenic operation of SiPMs requires a switch from doped CsI/NaI to undoped ones, as light yields of the latter operated at 77~K are about twice higher than those of the former at 300~K~\cite{Bonanomi52, Hahn53a, Hahn53, Sciver56, Beghian58, Sciver60, Fontana68, West70, Fontana70, Emkey76, persyk80, Woody90, Williams90, Wear96, Amsler02, Moszynski03, Moszynski03a, Moszynski05, Moszynski09, Sibczynski10, Sibczynski12, mikhailik15, csi16, ess19} (Note that cooling down existing doped crystals only provides mild increase of their light yields at around 230 K~\cite{COSINE21a}). The authors recently achieved a yield of $\sim$33.5~PE/keV$_\text{ee}$ using undoped CsI coupled to a Hamamatsu R11065 PMT with a peak quantum efficiency (QE) of $\sim$27\% at 77~K~\cite{ding20e}. A light yield of 40 to 50 PE/keV$_\text{ee}$ is achievable by switching from PMTs to SiPMs, which normally have a peak photon detection efficiency (PDE) of 40 to 50\%~\cite{jac14}. As for the second limiting factor, it has been measured that afterglow rates of undoped CsI at 77~K are compatible to CsI(Na) at room temperature~\cite{lewis21}, and the latter has been used by the COHERENT collaboration to observe coherent elastic neutrino-nucleus scatterings (CEvNS)~\cite{coherent17}. 

Reported in this paper is a measurement of the light yield of a detector system that consists of an undoped CsI crystal directly coupled to two SensL SiPMs at 77~K using $X$ and $\gamma$-ray peaks from an $^{241}$Am source. The operation of such a combination at 77~K was the first attempt in the world. 

\section{Experimental setup}
\label{s:exp}
\begin{figure}[htbp]
  \includegraphics[width=\linewidth]{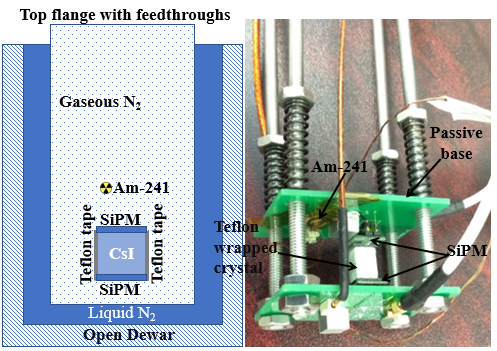}
  \caption{A sketch and a picture of the experimental setup.}
  \label{f:setup}
\end{figure}

\begin{figure}[htbp]
  \includegraphics[width=\linewidth]{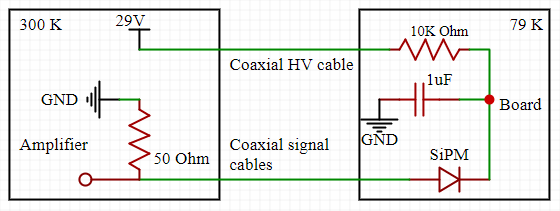}
  \caption{Circuit diagram of the passive base (right) and its wiring to room-temperature devices (left).}
  \label{f:circuit}
\end{figure}

\begin{figure}[htbp]
  \includegraphics[width=0.485\linewidth]{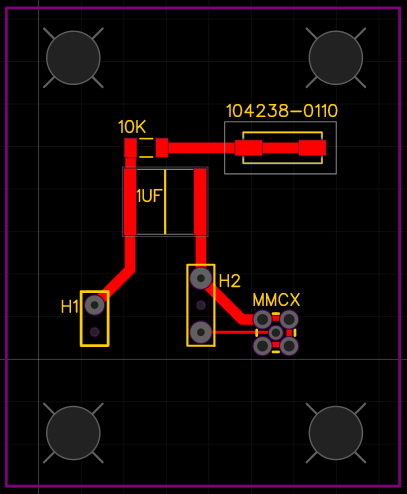}
    \includegraphics[width=0.495\linewidth]{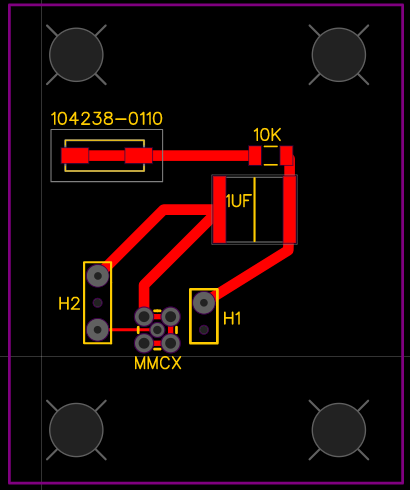}
  \caption{PCB layouts of the top (left) and the bottom (right) passive bases.}
  \label{f:bases}
\end{figure}

The experimental setup for the measurement is shown in Fig.~\ref{f:setup}. The undoped cuboid crystal was purchased from AMCRYS~\cite{amcrys}, which is 6~mm in length, width and 10~mm in height. All surfaces were mirror polished. To make sure there is no light leak, side surfaces of the crystal were wrapped with multiple layers of Teflon tape. Two MicroFJ-SMTPA-60035 (sensor size: 6$\times$6~mm$^2$, pixel size: $35 \times 35 \mu$m, total number of pixels: 18980) SiPMs from SensL~\cite{sipmJ} were used. A bias voltage of 29~V was applied to both sensors and was kept the same throughout the measurement. Breakdown voltages of these sensors were measured by SensL to be around 24.2$\sim$24.7 V at room temperature and were observed to change linearly with temperature at least down to -20$^\circ$C. The real over-voltage should have hence been higher than $29-24=5$ V. The two sensors were soldered on two pin adapter boards separately. Each pin adapter board was then inserted into a home-designed PCB, which only contained passive components, hence, was called a passive base. Fig.~\ref{f:circuit} shows the circuit diagram~\cite{diagram} of the base. The 1~$\mu$F capacitor was used to sustain large current during the avalanche (Geiger discharge) of a SiPM. It also forms a noise filter together with the 10~K$\Omega$ quenching resistor.  The current signal was converted to a voltage one on the 50 $\Omega$ internal impedance of the voltage amplifier used in this measurement. PCB layouts of the top and bottom passive bases are shown in Fig.~\ref{f:bases}. To ensure adequate optical contact without optical grease, the PCBs were pushed against two crystal end surfaces by springs. An $^{241}$Am source was attached on the top passive base for energy calibration.

To minimize exposure of the undoped CsI to atmospheric moisture, the assembly was done in a glove bag flushed with dry nitrogen gas. The relative humidity was kept below 5\% at 22$^{\circ}$C during the assembly process. The SiPM-crystal assembly was lowered into a stainless steel chamber from its top opening as shown in the left sketch of Fig.~\ref{f:setup}; the inner diameter of the chamber was $\sim 10$~cm, and the length is 50~cm long. The chamber was vacuum sealed on both ends by two 6-inch ConFlat (CF) flanges. The bottom flange was blank and attached to the chamber with a copper gasket in between. The top flange was attached to the chamber with a fluorocarbon CF gasket in between for multiple operations. Vacuum welded to the top flange were five BNC, two SHV, one 19-pin electronic feedthroughs and two 1/4-inch VCR connectors.

After all cables were fixed inside the chamber, the top flange was closed. The chamber was then pumped with a Pfeiffer Vacuum HiCube 80 Eco to $\sim 1.2\times {10}^{-4}$~mbar. Afterward, it was refilled with dry nitrogen gas to $\sim 1.8$ Kgf/cm$^2$ and placed inside an open liquid nitrogen (LN$_2$) dewar. The dewar was then filled with LN$_2$ to cool the chamber and everything inside. After cooling, the chamber pressure was reduced to slightly above the atmospheric pressure.

A few Heraeus C~220 platinum resistance temperature sensors were used to monitor the cooling process. They were attached to the side surface of the crystal, the top passive board, and the top flange to obtain the temperature profile of the long chamber. A Raspberry Pi 2 computer with custom software~\cite{cravis} was used to read out the sensors. The cooling process took about 20 minutes due to the small size of the crystal. Most measurements, however, were taken after about 40 minutes of waiting to let the system reach thermal equilibrium. The temperature of the crystal was -195.7 $\pm$ 0.3 $^\circ$C during measurements, which was almost the same as the LN$_2$ temperature.

The passive boards were powered by a RIGOL DP821A DC power supply~\cite{dp800}. A voltage of 29~V was applied to the SiPMs. According to their manuals, the PDE at this voltage is $\sim 50\%$ for MicroFJ-SMTPA-60035 at 420~nm. Signals were further amplified by a Phillips Scientific Quad Bipolar Amplifier Model 771, which has four channels, each has a gain of ten. Chaining two channels together, a maximum gain of 10$\times$10 can be achieved. A gain of twenty (10 $\times$ 2) was used. Pulses from the amplifier were then fed into a CAEN DT5720 waveform digitizer, which had a 250~MHz sampling rate, a dynamic range of 2~V and a 12-bit resolution. WaveDump~\cite{wavedump}, a free software provided by CAEN, was used for data recording. The recorded binary data files were converted to CERN ROOT files for analysis~\cite{towards}. 

\section{Single PE response}
\begin{figure}[htbp]\centering
  \includegraphics[width=\linewidth]{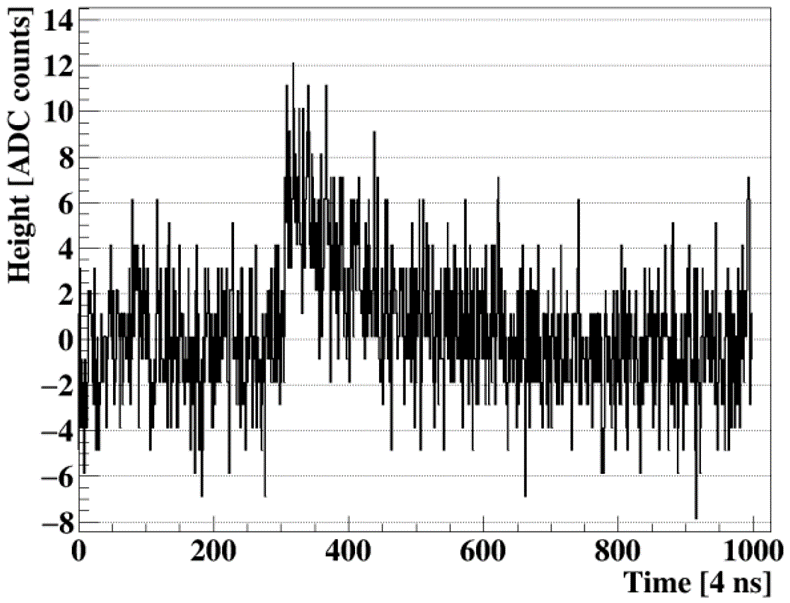}
  \caption{A random single PE waveform from the top SiPM. The ones from the bottom SiPM are very similar.}
  \label{f:single}
\end{figure}

Single PE responses of individual channels were investigated using waveform data triggered by dark counts. Some pre-traces were preserved before the rising edge of a pulse that triggered the digitizer to calculate the averaged baseline value of a waveform, which was then subtracted from each sample of the waveform. Fig.~\ref{f:single} shows a typical single PE pulse. The threshold was set by eye to keep a reasonable trigger rate ($\sim 8$~kHz for the top SiPM and $\sim 5$~kHz for the bottom one) while allowing the recording of most single PE pulses.

\begin{figure}[htbp]\centering
  \includegraphics[width=\linewidth]{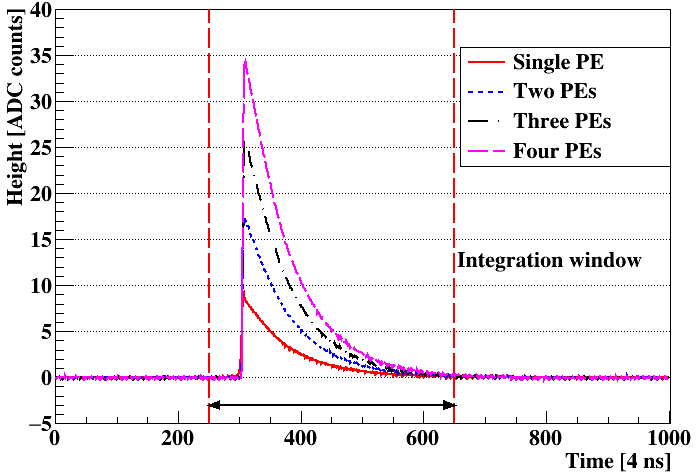}
  \caption{Average waveforms of different PEs from the top SiPM. The ones from the bottom SiPM are very similar.}
  \label{f:Ave}
\end{figure}

Overshooting or undershooting after a pulse may be hidden in a noisy baseline, especially for small PE pulses. A common way to remove the effect of electronic noise is to calculate the average waveform corresponding to the same PEs. For example, the average waveform of single PE was obtained by first adding up all single PE waveforms and then dividing the summed waveform by the total number of single PE events. The same method was used to obtain the average waveforms of higher PEs. They are all shown in Fig.~\ref{f:Ave}. No obvious overshooting or undershooting can be seen; and pulses of different PEs are well contained in the integration window. 

\begin{figure}[htbp]\centering
  \includegraphics[width=\linewidth]{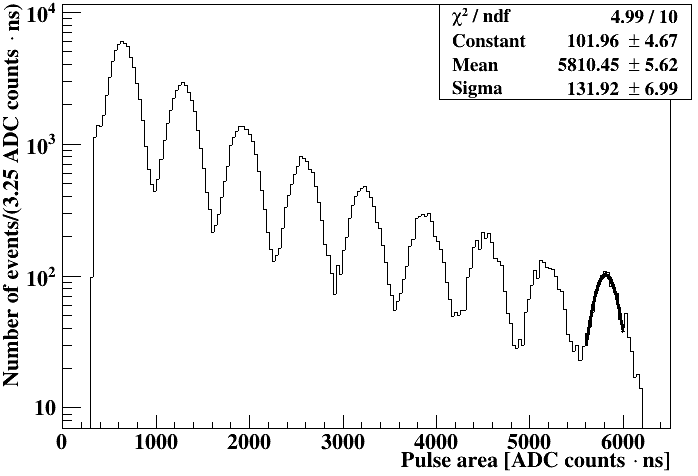}
  \caption{Single PE response of the top SiPM in logarithm scale. The ones from the bottom SiPM are very similar.}
  \label{f:1pe}
\end{figure}

Fig.~\ref{f:1pe} shows the distribution of pulse areas given by the integration, where individual PEs can be seen clearly. If the mean of the first peak is multiplied by $2, 3, 4, ...$ the results roughly match the means of the second, third, fourth, ... peaks. We hence believe that the first peak is the single PE distribution. The ninth peak was fitted using a Gaussian function to obtain its mean value and the result is shown in Fig.~\ref{f:1pe}. The same operations were done for all other peaks. The mean of single PE, mean$_\text{1PE}$, is defined as the Gaussian mean in Fig.~\ref{f:1pe} divided by the number of PEs, $n$. For example, the mean$_\text{1PE}$ for the ninth peak equals to $5810.45 / 9=645.61$ ADC counts$\cdot$ns.

\begin{figure}[htbp]\centering
  \includegraphics[width=\linewidth]{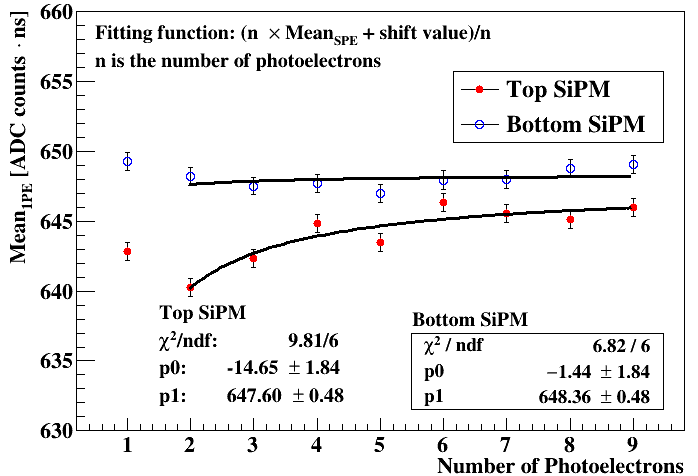}
  \caption{Mean$_\text{1PE}$ distributions obtained from top and bottom SiPMs.}
  \label{f:SPEfitting}
\end{figure}

The mean$_\text{1PE}$ as a function of the number of PEs is shown in Fig.~\ref{f:SPEfitting}. A flat line was expected while a slightly up-going curve was observed from the top SiPM. According to \cite{SiPM_fitting}, this is due to an overall shift of Fig.~\ref{f:1pe} to the left or right. To verify this idea, a function, as shown in Fig.~\ref{f:SPEfitting}, was fitted to the distribution, where, the mean$_\text{SPE}$ (p1) is the true mean of single PE before shifting, and the shift value (p0) is the amount of shift of the whole single PE response.  Note that the first points were not included in the fittings as some of the single PE pulses might not be able to pass the threshold, resulting in a slight distortion of the first peak in Fig.~\ref{f:1pe}.

According to the fitting, the means$_\text{SPE}$ (p1) of the top and bottom SiPMs are $647.60 \pm 0.48$~ADC counts$\cdot$ns and $648.36 \pm 0.48$~ADC counts$\cdot$ns respectively. The shift values (p0) are $-14.65 \pm 1.84$~ADC counts$\cdot$ns and $-1.44 \pm 1.84$~ADC counts$\cdot$ns respectively, which means that the true single PE response of the top SiPM was slightly shifted to the left, resulting in Fig.~\ref{f:1pe}. However, the origin of such a small shift is still unknown to the authors as the baseline has been removed prior to the integration. The same phenomenon has been observed in Ref.~\cite{SiPM_fitting}, the cause is also not explained.

\begin{figure}[htbp]\centering
  \includegraphics[width=\linewidth]{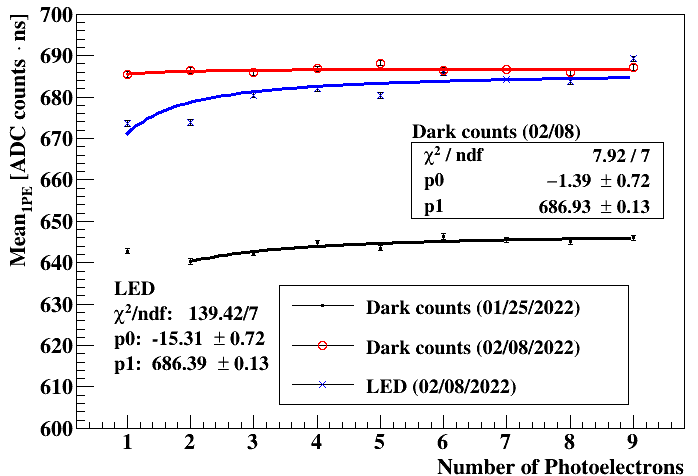}
  \includegraphics[width=\linewidth]{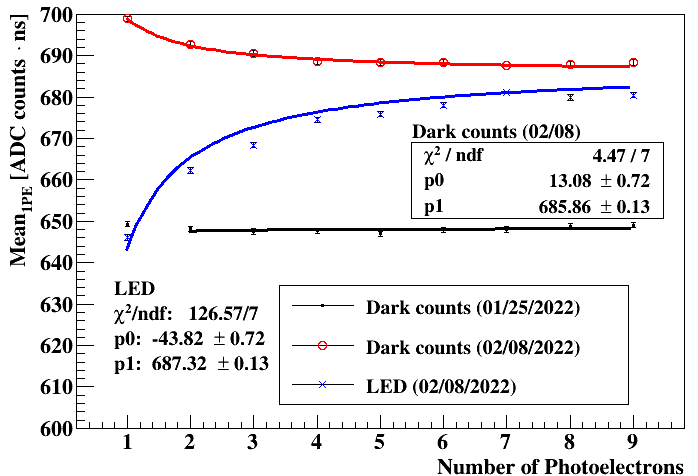}
  \caption{Mean$_\text{1PE}$ distributions of top SiPM (top) and bottom SiPM (bottom) obtained from different methods.}
  \label{f:compare}
\end{figure}

Single PE responses were also measured using an ultraviolet (370~nm) LED from Thorlabs. It was powered by a square pulse that last $\sim$50~ns and was emitted at a rate of 1~kHz from an RIGOL DG1022 arbitrary function generator. The voltage of the pulse was tuned to be around 4.55~V so that only zero or one photon hit the SiPM under study most of the time. Waveforms were recorded whenever a square pulse was generated. They were integrated in a fixed time window. The pulse area was plotted and fitted in a same way as described in the previous paragraphs. The results are compared with the dark count based ones and are shown in Fig.~\ref{f:compare}. Utilizing the same fitting process, the means$_\text{SPE}$ of the top and bottom SiPMs were found to be $686.39 \pm 0.13$~ADC counts$\cdot$ns and $687.32 \pm 0.13$~ADC counts$\cdot$ns. 

For fair comparison, another dark count based single PE measurement was done right after the LED measurement. The result is also included in Fig.~\ref{f:compare}. The means$_\text{SPE}$ from this measurement were $686.93 \pm 0.13$~ADC counts$\cdot$ns (top SiPM) and $685.86 \pm 0.13$~ADC counts$\cdot$ns (bottom SiPM). The later dark count based measurement gave slightly higher values than the earlier one. This indicates a possible gain shift of the SiPMs over time. 

The means$_\text{SPE}$ obtained in the earlier dark count based measurements (Fig.~\ref{f:SPEfitting}) were used for the light yield calculation (See Section~\ref{s:ly}), as they were done right after the energy calibration (See Section~\ref{s:ec}). The discrepancy of mean$_\text{SPE}$ in the earlier and later dark count based measurements was around 6.1\% for the top SiPM and 6.0\% for the bottom one. They are regarded as the uncertainties of mean$_\text{SPE}$. The results from the LED based measurements lay in between.

\section{Energy calibration}
\label{s:ec}
\begin{figure}[htbp]\centering
  \includegraphics[width=\linewidth]{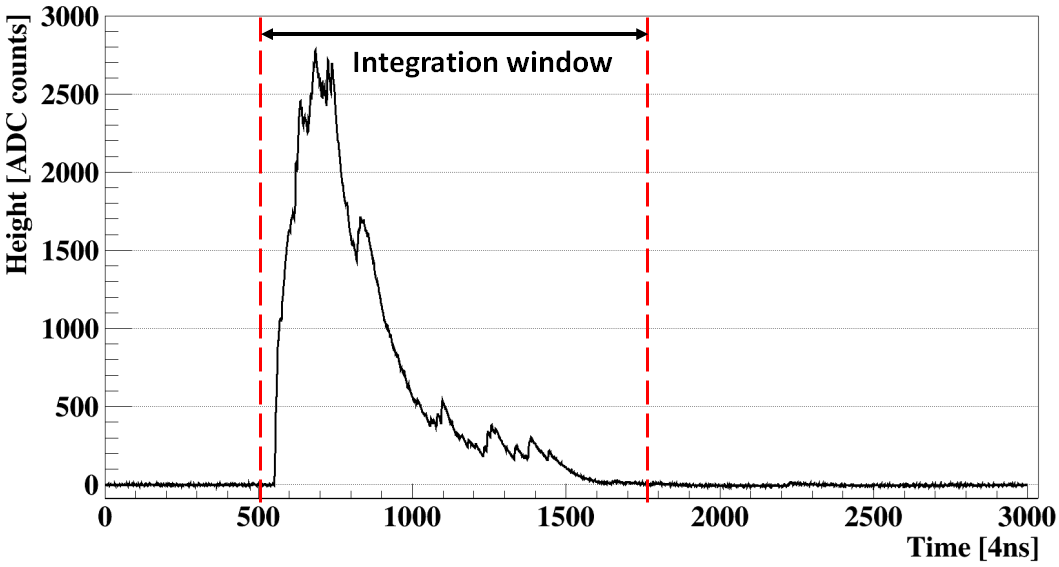}
  \caption{A randomly selected light pulse within the 59.5~keV peak from the top SiPM. The ones from the bottom SiPM are very similar.}
  \label{f:59.5}
\end{figure}

\begin{figure}[htbp]\centering
  \includegraphics[width=\linewidth]{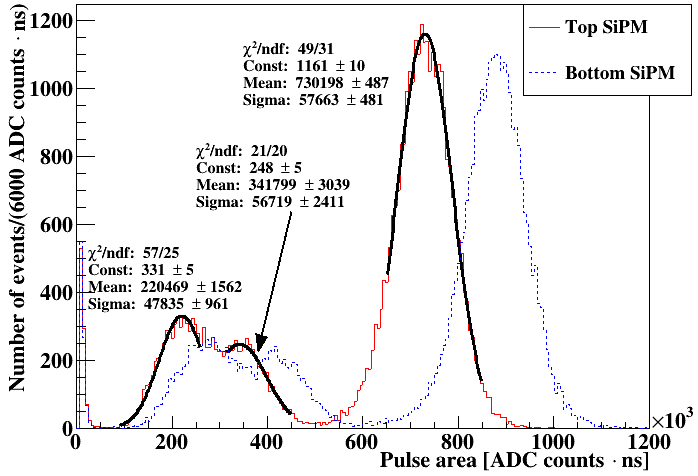}
  \caption{Energy spectrum of $^{241}$Am as the distribution of pulse areas in units of ADC counts$\cdot$ns.}
  \label{f:spec}
\end{figure}

The energy calibration was performed using $X$ and $\gamma$-rays from an $^{241}$Am radioactive source~\cite{ding20e, campbell86, toi}. The source was attached to the top passive board as shown in Fig.~\ref{f:setup}. The digitizer was triggered when heights of pulses from both SiPMs were more than 80 ADC counts. Pulses induced by radiation from the source were well above the threshold. The coincident trigger rate was around $\sim 100$~Hz. The integration of a pulse starts 50 samples before the trigger position and ends 10 samples after the position where the pulse goes back to zero. The integration window of a randomly selected light pulse induced by a 59.5~keV $\gamma$-ray is shown in Fig.~\ref{f:59.5}. The integration has a unit of ADC counts$\cdot$ns. The recorded energy spectrum in this unit is shown in Fig.~\ref{f:spec}.

The origin of each peak shown in Fig.~\ref{f:spec} was identified and summarized in Table~\ref{t:rPE}, based on Ref.~\cite{ding20e}, \cite{campbell86} and the \emph{Table of Radioactive Isotopes}~\cite{toi}. The peak around $200 \times 10^3$ ADC counts$\cdot$ns is a combination of multiple $X$-rays ranging from 13.8 to 20.1 keV. The averaged mean of these $X$-rays weighted by their intensities measured in \cite{ding20e} is 17.5~keV.
 
Peaks in Fig.~\ref{f:spec} were fitted with Gaussian functions to extract their mean values and widths. Most of the right side of the 17.5~keV peak and the left side of the 26.3~keV peak were excluded from the fitting as they overlapped with multiple X-ray peaks around 21~keV in between. The 59.5~keV peak was more or less Gaussian. However, its left side was slightly higher than the right side due to the loss of energy in materials between the source and the crystal~\cite{csi20}.

The bottom SiPM received more photons than the top one. This may be due to slightly different optical coupling conditions between the crystal and the two SiPMs. The top SiPM might not be full aligned with the top surface of the crystal. This is something that can be further improved. Another possible cause would be slightly different breakdown voltages of the two SiPMs. This would result in different over-voltages given the same bias, which causes different PDEs.

\begin{table}[htbp]
  \caption{\label{t:rPE} Fitting results of $^{241}$Am peaks in energy spectra for the top (top table) and the bottom (bottom table) SiPMs.}
  \begin{minipage}{\linewidth}\centering
  \begin{tabular}{cccccc}
    \toprule
    Type of & Energy & Mean$_\text{top}$ & Sigma & FWHM \\
   radiation & [keV$_\text{ee}$] & [ADC$\cdot$ns] & [ADC$\cdot$ns] & [\%] \\
    \midrule
     \begin{tabular}{r} $X$-ray\\ $\gamma$-ray \\ $\gamma$-ray \\
     \end{tabular} &
     \begin{tabular}{l} 17.5$^\dagger$ \\ 26.3 \\ 59.5 \\
     \end{tabular} &
     \begin{tabular}{r} 220469 \\ 341799 \\ 730198 \\
     \end{tabular} &
     \begin{tabular}{r} 47835.1 \\ 56719.3 \\ 57663.1
     \end{tabular} &
     \begin{tabular}{r} 51.1 \\ 39.1 \\ 18.6 \\
     \end{tabular} \\
     \bottomrule
     \end{tabular}
  \end{minipage}
  
  \begin{minipage}{\linewidth}\centering
  \begin{tabular}{cccccc}
    \toprule
    Type of & Energy & Mean$_\text{bottom}$ & Sigma & FWHM \\
   radiation & [keV$_\text{ee}$] & [ADC$\cdot$ns] & [ADC$\cdot$ns] & [\%] \\
    \midrule
     \begin{tabular}{r} $X$-ray\\ $\gamma$-ray \\ $\gamma$-ray \\
     \end{tabular} &
     \begin{tabular}{l} 17.5$^\dagger$ \\ 26.3 \\ 59.5 \\
     \end{tabular} &
     \begin{tabular}{r} 268360 \\ 413138 \\ 879406 \\
     \end{tabular} &
     \begin{tabular}{r} 56462.2 \\ 64188.3 \\ 61193.7 \\
     \end{tabular} &
     \begin{tabular}{r} 49.5 \\ 36.6 \\ 16.4 \\
     \end{tabular} \\
     \bottomrule
     \end{tabular}
  \end{minipage}
  $^\dagger$ Intensity averaged mean of $X$-rays near each other~\cite{ding20e}.\\
\end{table}

\begin{figure}[htbp]\centering
  \includegraphics[width=\linewidth]{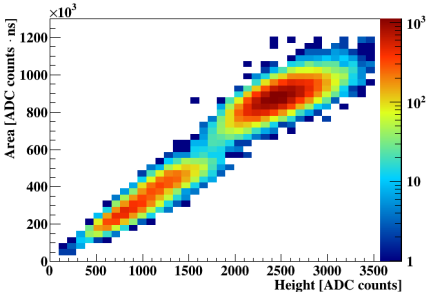}
  \caption{Pulse area versus pulse height for events collected with the $^{241}$Am source from the bottom SiPM.}
  \label{f:in}
\end{figure}

Fig.~\ref{f:in} shows the distribution of pulse areas versus pulse heights. A good linearity of the SiPM up to 59.5 keV was kept and the pulse height was controlled within the digitizer's dynamic range.

\section{Light yield}
\label{s:ly}
The fitted means of the 17.5 keV, 26.3 keV and 59.5 keV peaks in the $^{241}$Am spectrum in the unit of ADC counts$\cdot$ns were converted to the number of PE using the formula:
\begin{equation}
  \text{(number of PE)} = \frac{\text{(Mean-shift value) [ADC counts} \cdot \text{ns]}}{\text{mean}_\text{SPE}}.
  \label{e:m1pe}
\end{equation}

The shift value is added to account for the overall shift of the energy spectrum observed in the single PE measurement. However, compared to the mean value, it is much smaller, adding it to the equation does not change the final result much.

The light yield was calculated using the data in Table~\ref{t:rPE} and the following equation:
\begin{equation}
  \text{light yield }[\text{PE/keV}_\text{ee}] = \frac{\text{(number of PE)}}{\text{Energy }[\text{keV}_\text{ee}]}.
  \label{e:ly}
\end{equation}

\begin{figure}[htbp]\centering
  \includegraphics[width=\linewidth]{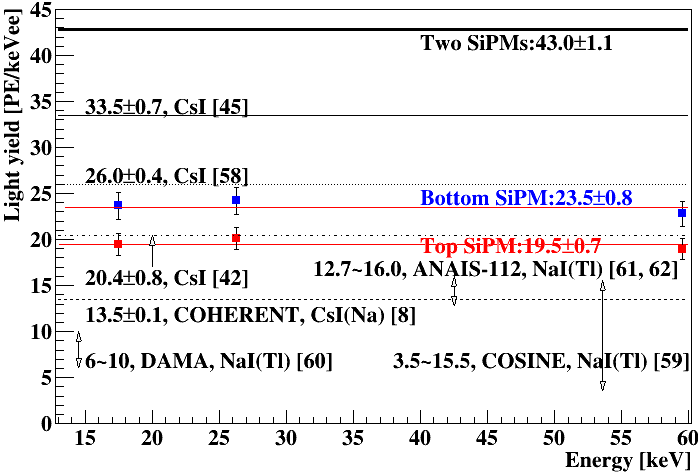}
  \caption{Light yields measured for scintillators made of iodide compounds from various experiments~\cite{cosine19,dama18,coherent17,csi16,csi20,ding20e,anais19, anais19a}. All were measured with PMTs except for the one in this work. The arrows only represent the range of light yield, which has no indication of energies of those experiments.}
  \label{f:ly}
\end{figure}

A flat line was fitted to the light yields obtained from the three peaks recorded in each SiPM, as shown in Fig.~\ref{f:ly}. The uncertainties of the light yield measurements are mostly determined by the uncertainties of the mean$_\text{SPE}$. The light yield observed by the top SiPM is 19.5 $\pm$ 0.7 PE$/$keV$_\text{ee}$, and by the bottom SiPM is 23.5 $\pm$ 0.8 PE$/$keV$_\text{ee}$. The total yield is hence 43.0 $\pm$ 1.1 PE$/$keV$_\text{ee}$. This and other results from related studies are compared in Fig.~\ref{f:ly}. Since sizes of crystals in these measurements are all different, this may not be a fair comparison if the self-absorption of optical photons in these crystals were significant. However, based on our limited experience, the influence of self-absorption is not as significant as those of optical surface conditions and light collection efficiencies of light sensors, as we observed larger light yields in larger crystals coupled to PMTs with higher quantum efficiencies and better wrapping of crystals~\cite{csi16,csi20,ding20e}.

\section{Energy resolution}
As seen in Fig.~\ref{f:ly}, the light yield obtained using SiPMs is higher than those obtained using PMTs. Assuming pure Poisson statistics, the energy resolution using SiPMs should be better than that using PMTs. However, this is not the case. The FWHM of the averaged 17.5~keV peak is $\sim$ 50\%, as shown in the last column of Table~\ref{t:rPE}. It may be explained by the merging of the 13.9, 17.8 and 21.0~keV peaks~\cite{ding20e}. The FWHM of the 26.3~keV and 59.5~keV peaks are $\sim$ 38\% and $\sim$ 18\%, respectively, while the PMT gives 28.5\% and 9.5\%, correspondingly~\cite{ding20e}. This discrepancy implies other contributions to the energy resolution in the SiPM setup, which have yet to be investigated.

The slightly worse energy resolution makes it difficult to resolve $X$-ray peak close to each other. However, it might not be a concern for low energy dark matter detection as the nuclear recoil spectrum has a shape close to an exponential decay near the threshold. The broadening of such a distribution does not necessarily reduce the number of observed events.

\section{Optical cross-talks between SiPMs}
As shown in Fig.~\ref{f:spec}, there is a small increase of event rate close to the threshold ($\sim15000$ ADC counts$\cdot$ns) from both SiPMs. However, there is no $X$-ray peaks around that region from the $^{241}$Am source. Certain instrumental noise might be the cause of this small bump, for example, optical cross-talks. However, for a cross-talk event to pass the two-SiPM coincident trigger, optical photons coming out of one SiPM must hit the other. Such a phenomenon is categorized as external cross-talk in Ref.~\cite{optical_crosstalks}.
\begin{figure}[htbp]\centering
  \includegraphics[width=0.36\linewidth]{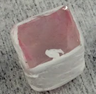}
    \includegraphics[width=0.57\linewidth]{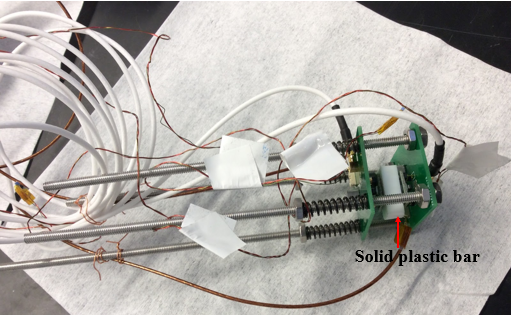}
  \caption{Left: transparent plastic cube. Right: opaque solid plastic bar in between two SiPMs.}
  \label{f:talk_set}
\end{figure}

To verify this possibility, several modifications of the experimental setup were done. First, the CsI crystal was replaced by a transparent hollow plastic cube with roughly the same dimensions, the picture of which can be seen in Fig.~\ref{f:talk_set}. Secondly, the radioactive source was removed. And finally, the plastic cube was replaced by an opaque solid plastic bar. Data were taken in coincident trigger mode with those modifications applied one by one. The coincident trigger rates are summarized in Table~\ref{t:optical}. 
\begin{table}[htbp] \centering
  \caption{\label{t:optical} Coincident trigger rates in different setups.}
  \begin{tabular}{cccc}
    \toprule
    &Experimental setups &Trigger rates [Hz]\\
    \midrule
     & CsI+$^{241}$Am & $750 \pm  50$\\
    \midrule
     & Transparent plastic cube+$^{241}$Am & $70 \pm 20$\\
    \midrule
     & Transparent plastic cube & $50 \pm 10$\\
     \midrule
     & Opaque solid plastic bar & $2 \pm 2$\\
    \bottomrule
  \end{tabular}
\end{table}

\begin{figure}[htbp]\centering
  \includegraphics[width=\linewidth]{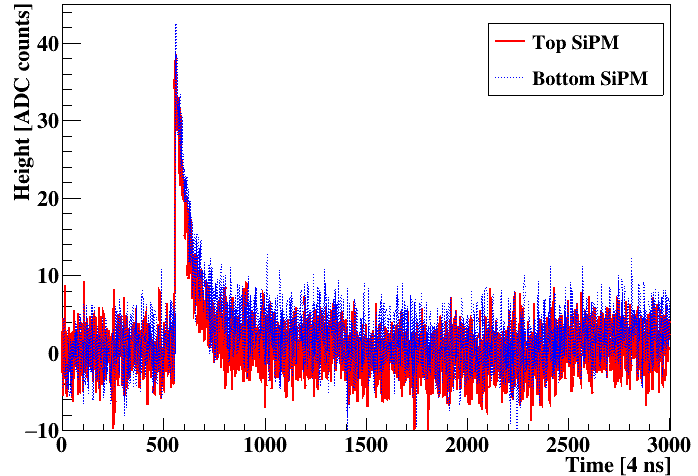}
  \caption{Waveforms from a common event taken with the transparent cube.}
  \label{f:talk}
\end{figure}

As shown clearly in Table~\ref{t:optical}, the trigger rate drops greatly when the scintillating crystal was removed. This is easy to understand, as most of the coincidentally triggered events were due to scintillation light from the crystal. It is troublesome to see that there were still quite some coincidentally triggered events after the source and the crystal were removed. They must come from the SiPMs. The time window for coincident trigger was set to be 8~ns. If light pulses in different SiPMs are due to random dark noise, their rising edges should appear randomly within 8~ns time window. Fig.~\ref{f:talk} shows waveforms from a common event taken with the transparent cube. Light pulses from two different SiPMs went across the threshold at exact the same time, indicating that they are highly correlated. The trigger rate dropped to nearly zero when the transparent cube was substituted by the opaque solid plastic bar. All these confirm that the events close to the threshold are indeed due to external optical cross-talks between the two SiPMs. 

\begin{figure}[htbp]\centering
  \includegraphics[width=\linewidth]{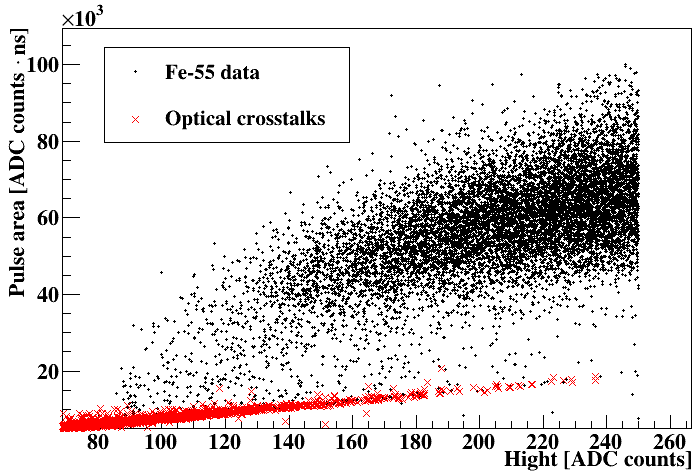}
  \caption{Area versus height of pulses in $^{55}$Fe (black dots) and optical cross-talks (red crosses) from the top SiPM.}
  \label{f:phy}
\end{figure}

One of the motivations to replace PMTs with SiPMs is to eliminate the Cherenkov radiation that would coincidentally trigger multiple PMTs. However, external optical cross-talks may coincidentally trigger multiple SiPMs as well. Some good methods to distinguish optical cross-talk events from physical ones are needed to justify the proposed replacement. A dedicated data set was taken with nothing in between two SiPMs to obtain optical cross-talk events. Another data set was taken with the CsI crystal and a $^{55}$Fe source to obtain low energy physical events. Fig.~\ref{f:phy} shows pulse area versus pulse height of waveforms in these data sets. The red crosses were from optical cross-talks, while the black dots were from $^{55}$Fe. Since pulses due to optical cross-talks are sharp and narrow (see Fig.~\ref{f:talk}), the area-to-height ratio is much smaller than that of physical events. Such a ratio can be a good parameter to remove events due to optical cross-talks.

Another method is to physically reduce the emission of optical photons from an avalanche cell, or to block them from reaching other cells~\cite{optical_crosstalks}. This is an active research area in the fabrication of SiPMs. Hopefully, this will become less a concern over time.

\section{Dark counts}
One major drawback of a SiPM array compared to a PMT is its high DCR at room temperature ($\sim$ hundred kHz). Fortunately, it drops quickly with temperature, and can be as low as 0.2~Hz/mm$^2$ below 77~K ~\cite{aalseth17}, while the PDE does not change much over temperature~\cite{oto07, lc08, aki09, jan11}. However, a SiPM array that has an active area similar to a 3-in PMT would still have an about 100~Hz DCR at 77~K. A simple toy MC reveals that a 10-ns coincident window between two such arrays coupled to the same crystal results in a trigger rate of about $10^{-5}$~Hz. This was also observed in this work that coincident trigger could dramatically reduce the trigger rate. If such a detector is placed near an accelerator-based neutrino source to detect neutrinos, a further time coincidence with beam pulses can be further requested to make the rate negligible.

\section{Readout electronics of SiPM arrays}
For neutrino or dark matter detection, crystals on the scale of $\sim10$~kg are needed, one surface area of which would be in the order of $10 \times 10$ cm$^2$. To fully cover such a surface, 400 SiPMs are needed assuming each has a surface area of $5 \times 5$ mm$^2$. The electronic readout of them could be a challenge. A natural option is to group the output of a few SiPMs into one channel. However, the total number of SiPMs that can be grouped is limited by the relatively large capacitance and DCR of individual SiPMs. Another possibility is to use CMOS-based ASIC to readout many channels with a single chip. CAEN has developed such a front-end system. It features a standalone unit, A5202, that contains 2 WEEROC CITIROC chips, each provides a multiplexed output of 32 SiPM channels. It also features a flexible micro coaxial cable bundle, A5260B, connecting a remote SiPM arrays with the A5202. Excellent single PE resolutions can be achieved with a cable length up to 3 m. The measurement mentioned in this work also confirmed that long cables still allow the observation of single PE pulses. Given the compact size of a scintillating crystal based detector, that length is sufficient to bridge cold SiPM arrays and warm ASIC front-ends. Warm electronics are more convenient to maintain, also reduce heating inside the cryostat, and will be tried out in our future work.

\section{Conclusion}
The light yield of an undoped CsI crystal coupled to two SiPMs at 77~K was measured to be $43.0 \pm 1.1$~PE/keV$_\text{ee}$ using $X$ and $\gamma$-ray peaks from an $^{241}$Am radioactive source. Assuming coincident trigger of two photons in two different SiPMs and a constant nuclear quenching factor of 5\%~\cite{lewis21}, the energy threshold of such a detector can be as low as 0.9 keV for nuclear recoils. To authors' knowledge, this is the first trial of such a combination in the world. The high light yield and the removal of Cherenkov light originally from PMT windows make it an attractive method to lower energy thresholds of inorganic scintillating crystal based detectors for dark matter and neutrino detection. Potential drawbacks of this novel combination compared to the traditional PMT readout at room temperature were identified, such as worse energy resolution, optical cross-talks between SiPMs, higher dark count rates, and many more readout channels, etc. Possible solutions and directions of further investigation were provided.

\begin{acknowledgements}
  This work is supported by the Department of Energy (DOE), USA, award DE-SC0022167, and the National Science Foundation (NSF), USA, award PHY-1506036. Computations supporting this project were performed on High Performance Computing systems at the University of South Dakota, funded by NSF award OAC-1626516.
\end{acknowledgements}

\bibliographystyle{spphys} % APS-like style for physics
\bibliography{ref}

\end{document}